\pdfoutput=1
\documentclass[runningheads]{llncs}
\usepackage{graphicx}
\usepackage{subfig}
\usepackage{bbm}
\usepackage{listings,chngcntr}
    \usepackage{xcolor}
\begin{document}

\title{Improving the Performance and Resilience of MPI Parallel Jobs with Topology and Fault-Aware Process Placement}

\author{Ioannis Vardas \and
Manolis Ploumidis \and
Manolis Marazakis}

\institute{Institute of Computer Science (ICS),
\email{\{vardas, ploumid, maraz\}@ics.forth.gr}}

\maketitle
\begin{abstract}
HPC systems keep growing in size to meet the ever-increasing demand for performance and computational
resources. Apart from increased performance, large scale systems face two challenges that hinder
further growth: energy efficiency and resiliency.  At the same time, applications seeking increased
performance rely on advanced parallelism for exploiting system resources, which leads to increased
pressure on system interconnects.  At large system scales, increased communication locality can be
beneficial both in terms of application performance and energy consumption. Towards this direction,
several studies focus on deriving a mapping of an application's processes to system nodes in a way
that communication cost is reduced. A common approach is to express both the application's communication
patterns and the system architecture as graphs and then solve the corresponding mapping problem.
Apart from communication cost, the completion time of a job can also be affected by node failures.
Node failures may result in job abortions, requiring job restarts.
In this paper, we address the problem of assigning processes to system resources with the goal of
reducing communication cost while also taking into account node failures. The proposed approach is
integrated into the Slurm resource manager.  Evaluation results show that, in scenarios where few nodes
have a low outage probability, the proposed process placement approach achieves a notable decrease in
the completion time of batches of MPI jobs. Compared to the default process placement approach in Slurm,
the reduction is 18.9\% and 31\%, respectively for two different MPI applications.
\end{abstract}
\section{Introduction}
There is a large number of applications, from a wide range of fields, such as, molecular dynamics, astronomy and
astrophysics, weather forecast and climate modeling, that address complex problems requiring a huge amount of
computational resources. Such applications rely on parallel machines and programming models that allow
harvesting their resources. The MPI standard \cite{ref_mpi_standard} is the dominant realization of such a
parallel programming model.

There is a wide range of different approaches for improving the performance of MPI applications including
topology or machine-aware collective primitives \cite{1419923,MA20131000,Sack:2012:FTC:2145816.2145823},
hardware assistance for certain MPI primitives \cite{6044798,Almasi:2005:OMC:1088149.1088183,1302912}
and point-to-point primitives, tuned for RDMA-capable
networks~\cite{4556083,Liu:2004:HPR:1119508.1119510,4536262}.
One approach that has received significant attention tackles the following problem: given an application with
multiple processes that exhibit a specific communication pattern and a parallel machine with several nodes, how should
processes be laid on the available resources so that some criteria are optimized. The MPI standard offers capabilities for exploiting topology related information, however,
leveraging such information for optimizing communications is either delegated to the MPI implementation, or the
resource manager.

A common approach for addressing the aforementioned process placement problem, is to assign system resources to processes
with a joint goal of minimizing communication cost while achieving a fair load balance among system nodes.
Assigning processes with a \textit{heavy} communication profile at \textit{nearby} nodes is expected to improve
overall application running time.
Improving communication locality has also been explored in the context of other performance criteria.
Power consumption of HPC systems has been pinpointed as an obstacle towards Exascale
\cite{daroa_ipto_2008}. Suitable mapping of processes to system resources can increase communication locality which
can in turn, reduce energy consumption due to
the interconnect \cite{Hoefler_ics_2011,daroa_ipto_2008}. The work in \cite{6507513,Hoefler_ics_2011} also
explores the potential to reduce network congestion.

The main contribution of this work is a process placement approach that aims at improving two aspects
of MPI job completion time.
The first goal is to reduce the communication cost incurred due to inter-node traffic.
The second goal of the proposed process placement approach is, to reduce the overhead
of a job abort due to node failures. Note that, a node failure may result in a call to an MPI
primitive returning an error. The default handling, provisioned by the
MPI standard, for such cases, is abortion of the job \cite{ref_mpi_standard}.
The cost of a job being aborted during execution becomes even more profound for codes that solve complex problems
and may require running for days.

Several studies have outlined the effect of system failures on resource utilization. Authors in \cite{darpa2009} report  that in a large scale HPC system, 20\% or more of the
computing resources are wasted  due  to  failures and recoveries. For one of Google's multipurpose clusters, it was found that a large fraction of time is spent for jobs that do not complete successfully~\cite{7980073}. In \cite{7266835} authors show that system related errors cause an application to  fail once every 15 minutes. What is more, failed applications, although few in number, account for approximately 9\% of total production hours. Authors in \cite{bianca-10.1088/1742-6596/78/1/012022} examine node failure rate in the dataset collected during 1995–2005 at LANL. The number of failures per year per system can be as high as 1100 meaning that an application requiring the full cluster is expected to fail more than two times per day.

For reducing the
communication cost of an MPI job, the proposed approach aims at
placing ranks with a heavy communication profile at nearby nodes (with
respect to topological distance). This assumes that a training run of
the MPI job is performed that has allowed to extract that job's
communication profile. For extracting the topological distance between
any two nodes of the parallel machine, topology information is also assumed to
be available. The corresponding topology mapping problem is solved using the Scotch graph mapping library
\cite{scotch_paper,ref_url_scotch}.

The second contribution is the integration of the proposed process placement approach into \textit{Slurm}.
\textit{Slurm} is
the resource manager and job scheduler for about $60\%$ of the machines in the Top 500 list
\cite{ref_url_top500}. For the evaluation of the proposed approach, batches of different MPI jobs where
simulated in the SimGrid \cite{casanova:hal-01017319} environment, which is a
distributed computer system simulator.
Simulated scenarios reveal that our placement approach achieves a notable decrease in batch completion
time, when compared to the default placement approach of Slurm. In scenarios where approximately $3\%$ of the nodes
exhibited an outage probability of $2\%$, the corresponding improvement over Slurm's policy was $18.9\%$ and $31\%$,
respectively, for the two MPI applications tried. Additionally, the proposed approach manages to reduce the job
instances that are aborted due to node failures.

The rest of the work is organized as follows: Section~\ref{related_work} provides a more detailed discussion
of related studies, while Section~\ref{tofa_approach} presents the proposed process placement approach.
In Section~\ref{sec_tofa_slurm_ingegration}, details about the integration of
the proposed process placement approach in Slurm are provided. Evaluation results are
presented in Section~\ref{sec_evaluation} while
conclusions and future work are discussed in Section~\ref{sec_conc_future_work}.

\section{Related work} \label{related_work}

There is a wide range of different approaches, concerning different stack layers, for improving the performance achieved
by MPI applications. There are studies related to hardware, systems software like resource managers,
the MPI library and the application itself. A taxonomy of related studies can be derived by
considering the type of feedback they use, that is, feedback related to the system architecture,
or topology and information regarding the communication pattern or profile of a specific job.

Example of studies that use neither topology, nor application communication profile related feedback are those that
focus on optimizing collective or point-to-point primitives.
Work in \cite{Ruefenacht:2017:GRD:3163938.3164013} for example, suggests a new implementation for the Allreduce primitive,
that achieves reduced execution time over the well known recursive doubling algorithm.
Several studies, focus on optimizing or tuning, the messaging protocol employed by point-to-point
primitives, for the case where
RDMA capabilities are offered by the platform \cite{4556083,Liu:2004:HPR:1119508.1119510,4536262}.
Work in \cite{6044798,Almasi:2005:OMC:1088149.1088183,1302912}, either introduces or, utilizes hardware support for
speeding up certain MPI primitives.

There are approaches that utilize knowledge regarding the topology or system architecture.
Such approaches focus on deriving topology aware, or, improved implementation of collective primitives for specific
platforms \cite{1419923,MA20131000,Sack:2012:FTC:2145816.2145823}.
Authors in \cite{Sack:2012:FTC:2145816.2145823} for example, suggest non-minimal algorithms for allgather and
reduce-scatter primitives for the case of Clos, single-, and multi-ported torus networks.

Studies that are mostly related to our work, employ as input a graph that captures a problem's communication patterns
along with a graph that models the architecture of the platform available.
Most of them address the mapping problem described in~\cite{1675756}, which seeks to assign processes of an application
onto computational resources of the available platform.
Authors in \cite{Hoefler_ics_2011} for example, address the problem of mapping arbitrary communication
topologies to arbitrary-heterogeneous
machine topologies with the goal of minimizing the maximum congestion and average dilation.
The work in \cite{7546612} suggests a new algorithm for reordering ranks of an MPI jobs so that communication
cost is minimized.
In \cite{rodgri_iscc_2009}, the dual recursive bi-partitioning method is explored for solving the graph
mapping problem, for the case of clusters with multi-core processors.
Authors in \cite{1592864} show that, for the case of hierarchical systems especially, the topology mapping problem
can be solved through weighted graph partitioning. They discuss the properties of a new Kernighan-Li heuristic
for providing direct graph k-partitioning.
Authors in \cite{6495451} explore an enhanced version, in terms of complexity, of the TreeMatch algorithm
\cite{10.1007/978-3-642-15291-7_20} for deriving an optimized process placement on the platform available.
The work in \cite{6061055} discusses an efficient algorithm based on the Kernighan-Li heuristic for assigning tasks
of a parallel program to processors for the case of hyper-cube parallel computers.

A similar approach formulates the problem of assigning processes onto system resources as a \textit{Quadratic Assignment
Problem (QAP)} \cite{1675756}.
Authors in \cite{6507513} for example, formulate the assignment of processes to nodes as a QAP. They suggest a
heuristic based on graph partitioning and the greedy randomized adaptive search for solving the corresponding problem.

The approaches discussed so far are profile guided, that is, they assume that the underlying communication pattern of
a specific application is available. There are approaches though, that do not carry this dependence.
Authors in \cite{7530080} for example, explore four heuristics, to perform rank reordering for realizing run-time
topology awareness, for the case of the MPI Allgather primitive. The corresponding approach does not rely on an
application's profile. Instead, it is based on the communication pattern exhibited by each algorithm used
in the allgather primitive. The work in \cite{6061150} targets mapping the logical communications implied by
broadcast tree to the physical network topology in order to reduce delays at critical phases of the
broadcast schedule.

The work in \cite{6061055} discusses an approach that is is complementary to that of assigning processes
of a parallel job onto platform resources. More precisely, a framework is suggested that
rearranges the logical communication of broadcast and allgather
operations taking into account process distances instead of process ranks along with topology information.

For mitigating the impact of failures, failure awareness has been integrated in several different approaches including
checkpointing \cite{6114447,1630866,6903564}, scheduling methods \cite{993209,4343845} and methods for resource
allocation and resource management \cite{10.1007/11407522_13,10.1016/j.jpdc.2010.01.002,5488431}.

\section{Proposed process placement approach} \label{tofa_approach}

In this work, we consider an approach for assigning process of an MPI job onto nodes of a given platform, named,
\textit{TOpology and Fault Aware (TOFA)} process placement approach.
We focus on MPI applications with a static profile, that is, applications where
processes coexist for the whole duration of the execution.
In accordance with relevant studies \cite{Hoefler_ics_2011,1592864},
we formulate the problem of assigning process to platform nodes as a
topology mapping problem. We model the communication among different process as an undirected graph $G=(V_{G}, E_{G})$.
For the rest of the study, this graph will be referred to as the \textit{communication graph}.
Each vertex $v_{g} \in V_{G}$ corresponds to one process while an edge $e \in E_{G}$ connecting vertices
$u_{g}$ and $v_{g}$ denotes communication between the corresponding processes. Edge weights may depict either number of
messages or total traffic exchanged between the two processes.
 As it stated in~\cite{8411068} the choice between volume and number of messages is not standard but rather application depended. Thus, each application should be tested before choosing the best way of depicting the edge weight of the communication graph. After testing both cases, the evaluation results presented in Section~\ref{sec_evaluation} are derived by considering total traffic volume as edge weights.

\begin{figure}
     \subfloat[]
     {
         \includegraphics[scale=.31]{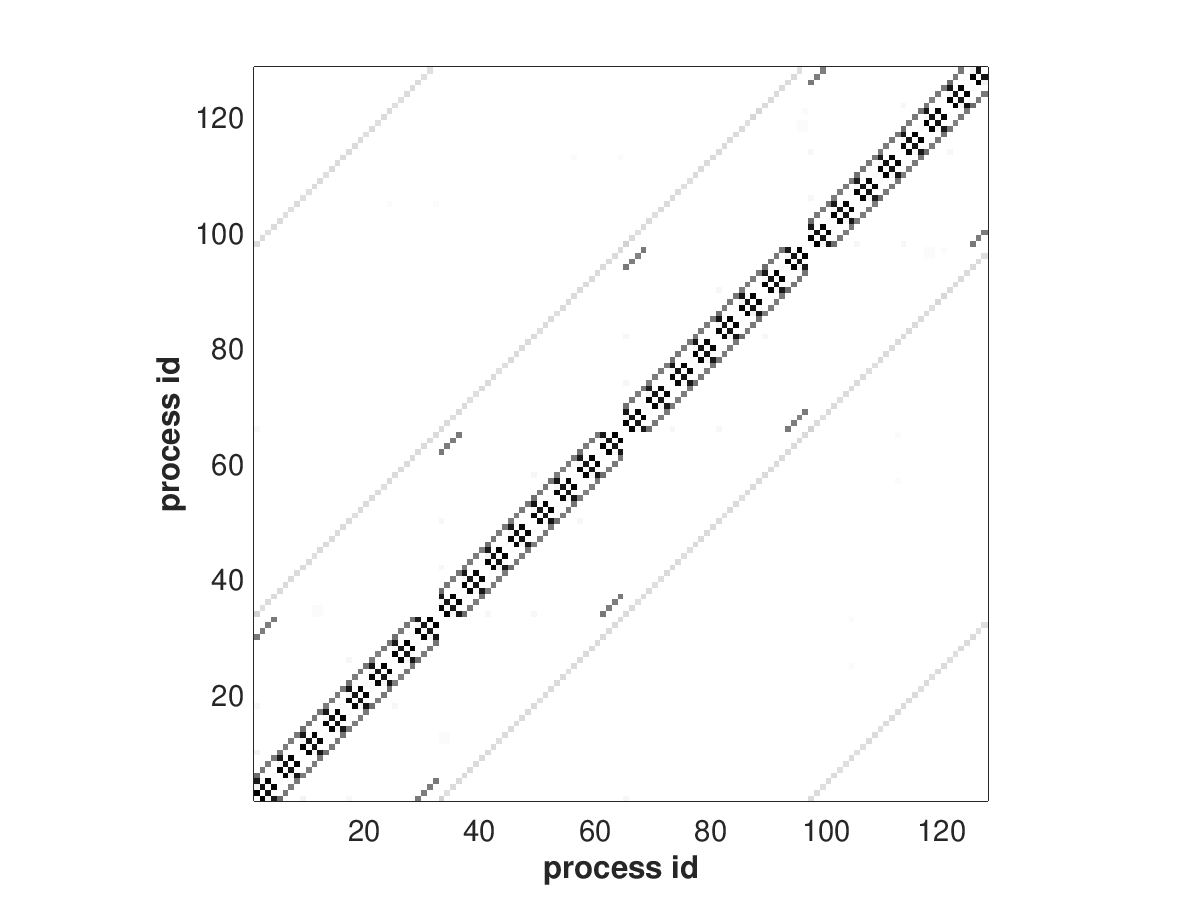}
         \label{fig_lammps_heatmap}
     }
     \subfloat[]
     {
         \includegraphics[scale=.32]{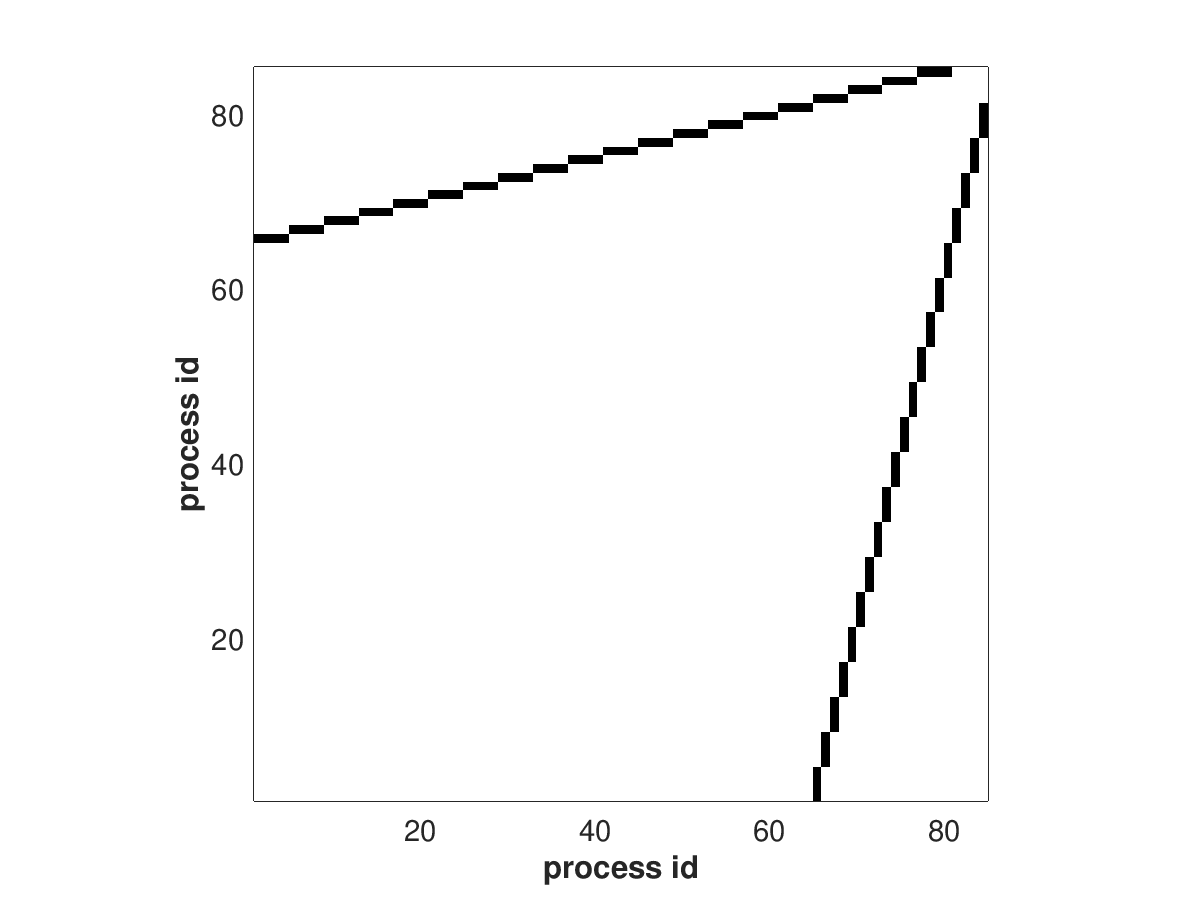}
         \label{fig_npb_dt_c_heatmap}
     }
     \caption{a)LAMMPS run with 128 processes, b)NPB-DT class C run with 85 processes}
\end{figure}

For extracting the number of messages, and total bytes exchanged between each pair of processes, we have implemented a
custom profiling tool for MPI applications. This tool is a dynamically linked library that intercepts all
calls to MPI primitives that initiate traffic, such as, point-to-point, collective, and one-sided ones.
The output produced consists of two graphs, namely, $G_{v}$ and $G_{m}$. Each of these graphs is
represented as a matrix of size $N\times N$, where $N$ is the number of
processes involved in the MPI program. Graph $G_{v}$ captures the
number of bytes exchanged for each pair of processes, while graph
$G_{m}$, captures the corresponding number of messages. For the case
of $G_{v}$ for example, element $G_{v}(i,j)$ captures the sum of the
bytes sent from MPI rank $i$ to rank $j$ and the bytes sent from $j$ to $i$.
For the case of
collective primitives, the profiling tool is tuned to emulate the appropriate
algorithm for each collective. In this way, it is able to accurately capture the
traffic exchanged between each pair of processes during each phase of
that collective's schedule. Graph $G_{v}$, or $G_{m}$, is used to denote the guest graph $G$, mentioned in the previous
paragraph.
An additional feature of this profiling tool is that it records traffic through communicators
other than the default one. For correctly updating $G_{v}$ and
$G_{m}$, the rank of a process in a communicator other than
\textit{MPI\_COMM\_WORLD}, is transformed to the rank in
\textit{MPI\_COMM\_WORLD} ($R_{comm\_world}$), and counters in $G_{v}$
and $G_{m}$ are updated based on $R_{comm\_world}$.
Another feature of our profiling tool is that it produces a traffic heatmap, which depicts the amount of bytes
exchanged between each process pair. This traffic heatmap allows for visual inspection of the
corresponding application's communication pattern. This inspection offers insight about regularity or
irregularity of that pattern.
Figures \ref{fig_lammps_heatmap} and \ref{fig_npb_dt_c_heatmap}, depict the traffic heatmap produced by our profiling
tool, for the case of a LAMMPS \cite{ref_url_lammps_bench} run and the DT NAS parallel benchmark (NPB)
\cite{Bailey91thenas} respectively. The darker the data point, the higher the amount of traffic exchanged for the
corresponding process pair.
As Figure \ref{fig_lammps_heatmap}  shows, LAMMPS exhibits a more regular communication pattern with traffic points
being close to the main diagonal on a similar manner for all processes.

It should be noted that, the proposed process placement approach is profile-guided. This means that, for deriving an
assignment of MPI processes to nodes, a training run should be carried first, to derive the corresponding communication
graph. However, this cost can be amortized by performing multiple runs of the same application using the same input or
configuration. It should be noted though that, the overhead and the dependence on the application profile can be
avoided by utilizing the \textit{virtual topology} provisioned by the MPI standard. Virtual topologies can be adopted
by MPI applications to represent the communication pattern of that application. Nodes in that virtual topology correspond
to processes and edges connect processes that communicate.

The underlying platform is modeled through topology graph $H=(V_{H}, E_{H})$. Each vertex $v_{h} \in V_{H}$
corresponds to one node. Vertices carry no weight.
Let $R$ denote the routing logic of the underlying platform.
Assuming a 3D torus topology with fixed routing, the routing function $R(u,v)$
provides the list of links to be traversed by a message sent from node $u$ to $v$.
The weight $w(e_{u,v})$ of $e(u,v) \in E_{H}$ is set to be the number of hops traversed to reach $v$ when
starting from $v$.

A key characteristic of the proposed process placement approach from
is that, assignment of processes to nodes also takes into account node failures.
The fault model assumed is the following; nodes may fail independently of each other.
With the term \textit{failure}, we refer to any hardware- or software-related event, or reboot, that may constitute
the node temporarily unavailable. We further assume that a node restart is enough to fix transient failures
and that restart takes place instantaneously, i.e. we disregard recovery time.
When a node enters the failed state, it is incapable of performing both computation and communication, i.e.
cannot send, receive, or forward traffic on behalf of other nodes. Consequently, communication attempts
initiated by the MPI library will result in error and, in turn, job abortion. Moreover, when a node is in the
failed state, it is not able to respond to probes (\textit{heartbeats}) aimed at inferring its availability.
In this way, the corresponding
Slurm module that collects heartbeats is able to infer node outage.
Finally, we assume that there is no checkpoint/restart mechanism; thus, after a node fails any affected application
restarts from scratch.
Details regarding the heartbeat mechanism will be deferred until Section~\ref{sec_tofa_slurm_ingegration}.

Avoiding nodes that fail frequently is expected to decrease
the probability of a job being aborted. The importance of node failures becomes even more prominent for applications
that have large running times. However, a strict policy of avoiding failed nodes may force selecting a more sparse
subset of available nodes. This in turn may have an adverse effect on communication cost. There is thus a trade-off,
between the abort ratio tolerated and the communication cost. Therefore, a placement approach with
some tolerance for node failures may strike a more favourable balance between abort ratio and completion time for a batch
of jobs.

With TOFA, we approximate the effect of node failures on the cost of traversing a path by assigning
larger weights to paths that include nodes with a non-zero outage probability. Our initial choice of
small increases in the path cost revealed that this approach achieved only a marginal decrease in the probability
of aborting a job.
For that reason, we changed our approach for updating the cost of traversing paths with failing nodes as follows.
Assume that outage probability is available for each node.
From the routing function $R(u, v)$ we infer the list of links to be traversed by a message
sent from $u \in V_{H}$ to $v \in V_{H}$, where $H$ corresponds to the topology graph.
For each link $l \in R(u, v)$, $l^{s}$ and $l^{d}$ denote the origin and target of that link respectively.
Using this information, we maintain a registry, where input is a node id and output
is the list of paths with this node serving as an intermediate hop. Combining information provided by
the routing function and node outage probabilities, we update edge weights in the topology graph
using Equation~\ref{eq_edge_weights_crude}.
\begin{equation}\label{eq_edge_weights_crude}
w(e_{u,v}) = \sum_{l \in R(u,v)} c + c \times 100 \times \mathbbm{1}[ (p_{l^{s}}^{f}>0) \lor (p_{l^{d}}^{f}>0)],
\end{equation}
In Equation~\ref{eq_edge_weights_crude}
$p_{l^{s}}^{f}$ and $p_{l^{d}}^{f}$ are the failure probabilities
for nodes $l^{s}$ and $l^{d}$ respectively, and the constant $c$ denotes the cost in terms of number of hops.
Moreover, $\mathbbm{1}[p_{l^{s}}^{f}>0]$ is an indicator function with a value of 1 if $p_{l^{s}}^{f}>0$.
Our rationale is that if either of the two nodes involved in a link
has an outage probability other than zero, then the cost of that
link is set to $100$ instead of one. Thus, the cost of a failed
path becomes significantly higher than the cost of traversing
the longest path (in terms of number of hops) on the platform.

Having described how the communication and topology graphs are populated,
the final step is to describe the proposed process placement approach.
TOFA uses the Scotch graph mapping library \cite{scotch_paper,ref_url_scotch} to solve the
corresponding graph mapping problem. The output of TOFA is a mapping of vertices (processes) of communication graph G,
onto vertices on the topology graph H (corresponding to platform nodes).
Listing~\ref{lsting_tofa_pseudocode} describes in pseudocode TOFA's process placement steps.

\begin{tabular}{c}
    \begin{lstlisting}[caption={TOFA process placement approach},basicstyle=\ttfamily\scriptsize,mathescape=true,label={lsting_tofa_pseudocode}]
    Input: G
        Graph modeling an application's communication pattern
    Input: H
        Graph resembling the topology. Edge weights estimated through
        Equation 1 or 2
    Output: T
        T = <Process Id, Node Id>

    procedure TOFA(G, H):
        S=Find $|V_{G}|$ consecutive nodes s.t. $p_{n}^{f}=0, \; \forall n \in V_{H}$
        if $S = \lbrace \emptyset \rbrace$: then
            T := ScotchMap(G, H)
        else
            $H_{S}$ := ScotchExtract(H, S)
            T := ScotchMap(G, $H_{s}$)
        end if
        return T
    \end{lstlisting}
\end{tabular}

The input to the proposed process placement approach consists of the communication graph G
and the topology graph H.
For capturing the effect of node failures, edge weights in the topology graph are updated
according to Equation~\ref{eq_edge_weights_crude}.
The output of TOFA is set $T$, which has one entry per process. Each entry consists of two values, one corresponding to
the process id and a second one denoting the id of the node where that process is assigned.
In step 10, TOFA first searches for a set of consecutive nodes that
have zero outage probability. If this set is non-empty, then the set of $|V_{G}|$ fault-free nodes is passed as input
to Scotch library along with the topology graph $H$. Scotch then produces a new topology modeled through graph $H_{s}$
which is a subset of the original topology depicted through $H$. This functionality is denoted as \textit{ScotchExtract}
in step 14 of the above pseudocode. In step 15 then, Scotch is used to map the communication graph G in $H_{s}$.
If TOFA is unable to identify $|V_{G}|$ consecutive fault-free nodes, Scotch is used to map the communication graph into
the topology graph (step 12).

\section{Integration of TOFA in Slurm} \label{sec_tofa_slurm_ingegration}
In this section we discuss technical details regarding the modifications required to integrate TOFA into Slurm.
First a brief overview of Slurm is required.

\textit{Slurm} is the acronym for \textit{Simple Linux Utility and Resource
Manager}. It is the resource manager
and job scheduler for $60\%$ of the systems in the list of Top500 \cite{ref_url_top500}.
Slurm manages resources, such as, cores, CPUs, nodes, and memory. There is also support for the so-called \textit{generic
resources (GRESs)}, such as, Graphics Processing Units (GPUs), and CUDA Multi-Process Service (MPS).

Slurm's components can be broadly categorized in two sets. The first set contains the main daemons, that is, Slurm
controller and Slurm daemon denoted as \textit{slurmctld} and \textit{slurmd}
respectively. Slurmctld runs on a single node and is
responsible for allocating resources and scheduling jobs.
\textit{Slurmd} is the daemon
that runs on each compute node and waits to execute work issued by
\textit{slurmctld}. The second set contains user and administration tools like, \textit{srun}, \textit{sbatch},
\textit{squeue}, e.t.c. Srun and sbatch are used to initiate jobs while squeue is used to view information
about jobs located in the scheduling queue.

One powerful feature of Slurm is that its functionality
can be extended through plugins. The topology plugins for example, are used
to provide feedback regarding the system topology, enabling topology-optimized
resource selection. The resource selection plugins determine how resources are allocated to a job.
The generic resources plugin (GRES) can be used to manage non-standard or custom resources.
One plugin that was valuable for integrating TOFA into Slurm is \textit{SPANK}, which stands for
\textit{Slurm Plugin Architecture for Node and job (K)control}. It is a generic
interface for plugins which can be used to dynamically modify the job launch process.
\begin{figure}
  \centering
  \includegraphics[width=0.7\textwidth]{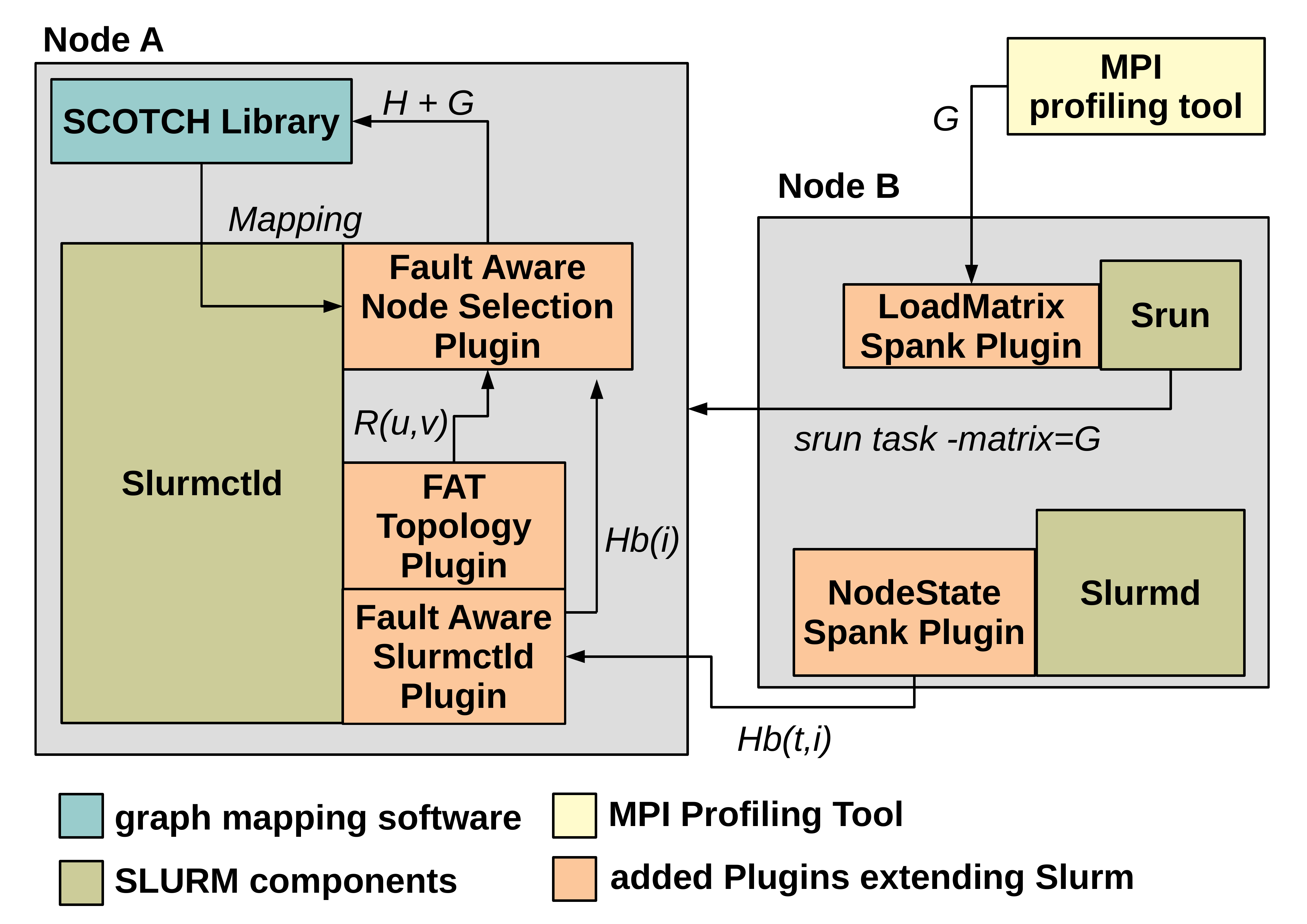}
  \caption{Plugins and components for integrating TOFA in Slurm} \label{fig_tofa_slurm}
\end{figure}

For integrating TOFA into Slurm, five different plugins were created along with  modifications of its source code. Our goal is to enable Slurm to utilize TOFA only when it is requested by the user while not interfering with the standard resource allocation path of Slurm.
The \textit{NodeState} and \textit{LoadMatrix}
SPANK plugins run in every compute node, whereas \textit{FAT
Topology}, \textit{Fault Aware Slurmctld} and \textit{Fault Aware
Node Selection} (FANS) plugins run only on the controller node.
Figure~\ref{fig_tofa_slurm} overviews the integration of TOFA into Slurm.

The \textit{Fault Aware Slurmctld} plugin, is responsible for periodic polling of each node through a heartbeat.
In Figure~\ref{fig_tofa_slurm}, the heartbeat of node $i$ at internal $t$ is denoted as $Hb(t,i)$.
Absence of a reply to a heartbeat is translated as node outage.
Slurmctld maintains a record of heartbeats for each node $i$, denoted as $HB(i)$.
Node outage probability can be inferred by post-processing
the history of each node's heartbeats.
Then, node outage probabilities can be combined with output from the routing function $R(u,v)$, according to Equation
\ref{eq_edge_weights_crude}, to update edge weights in the topology graph.
This plugin is also the context where different policies for inferring
node outage probability can be implemented. One such policy could be a moving or weighted moving average.

The \textit{NodeState} plugin, is a SPANK plugin located on each node, and is run once during the initialization
of the slurmd daemon. This plugin is responsible for replying to the heartbeats sent by the \textit{Fault Aware Slurmctld}
that is running on the controller node.

The \textit{LoadMatrix} plugin in Figure~\ref{fig_tofa_slurm}, is a SPANK plugin used to send the communication
graph $G$ from any compute node to the controller node.
Recall that, the communication graph is produced by the MPI profiling tool described in Section~\ref{tofa_approach}.
This plugin enables srun to have an extra argument which can be used to provide the
file containing a representation of $G$. Information regarding the communication graph $G$
will be sent to slurmctld where the actual assignment of processes to nodes will take place.

The \textit{Fault Aware Torus Topology (FATT)} plugin is responsible for implementing routing function $R(u,v)$.
Additionally, it provides a representation of the platform topology.
This representation is in the form of a graph and does not take node outage probabilities into account.
The creation of this graph takes place during slurmctld initialization.
This plugin reads a topology file which contains one entry for each node. This entry includes the id of the node
along with $x$, $y$, and $z$ coordinates on the 3D torus assumed.
Using this information, \textit{FATT} plugin realizes the routing function $R(u,v)$.
This function provides all intermediate nodes traversed for routing traffic from a node $u$ to a node $v$.
This information is required by the plugin \textit{Fault Aware Node Selection}, which is the plugin that performs
the actual allocation of resources.
Although Slurm offers a topology plugin suitable for 3D Torus, it cannot be utilized by our process placement approach
since it does not export routing related information similar to routing function $R(u,v)$.

The core functionality of resource selection is implemented by the \textit{Fault Aware Node Selection plugin}
which uses the following information as input:
\begin{itemize}
    \item Communication graph $G$ from the \textit{LoadMatrix} plugin
    \item Distance and intermediate nodes required for reaching node $v$ from $u$ and vice versa. This
        information is provided by the \textit{FATT} plugin
    \item Outage probability for each node which is the result of processing $Hb$ data structures at the
        \textit{Fault Aware Slurmctld} plugin
\end{itemize}

The main task performed by the aforementioned plugin is to integrate Scotch library's functionality.
More precisely, it invokes Scotch providing as input to it the application's communication graph $G$ and the
\textit{fault-aware} topology graph $H$.
Scotch in turn solves the actual graph mapping problem and returns as output an array with one entry per process.
Each entry has the id of the corresponding process and also the id of the node where that process is assigned. To ensure that each task will be executed on the node allocated through the aforementioned process, Slurm's default task layout process should be overiden. For achieving this, limited modifications to srun, sbatch and the Slurmtctld step manager were required. First, we added support for a new value for srun's "distribution" parameter. An srun command issued with `distribution=TOFA` and a file resembling the applications communication graph will enable Slurm to spawn each task on the node selected by our resource allocation apporach.

\section{Evaluation} \label{sec_evaluation}
In this section we present the evaluation of the proposed process placement approach.
As already stated, TOFA relies on Scotch for solving the corresponding graph mapping problem to
derive a layout of MPI processes on platform nodes.
The first part of the evaluation, in Section~\ref{sec_eval_part_a}, focuses on
assessing the quality (in terms of application performance)
of the mapping produced by Scotch. In that part, node failures are
disregarded, in other words, all nodes are assumed to have zero outage probability.
The evaluation of the proposed process placement approach is finally presented in Section~\ref{sec_eval_part_b}.

For the evaluation of TOFA we rely on the SimGrid \cite{casanova:hal-01017319}
framework for the simulation of applications that execute on distributed systems.
Within the SimGrid framework, the SMPI \cite{degomme:hal-01415484} interface is capable
of simulating unmodified MPI applications. In SMPI, the communication calls of the application are intercepted and
simulated, whereas the computations are carried out on the host machine.

Having a simulated environment for evaluation purposes offers several benefits.
The main one is that it offers a convenient way to emulate node failures.
Emulating or injecting faults in a real platform is more complicated.
Secondly, it allows to run multiple scenarios in parallel.
Finally, it allows to experiment easily with different topologies and explore their effect on the
process placement approach. With a real platform, this might not be so trivial.

For running MPI applications in the SimGrid environment, a description of the simulated platform is needed.
The main components of a simulated platform are: links between nodes, nodes
and routes. A node is characterized through a fixed computing
capability, in terms of \textit{floating point operations per second (FLOPS)}.
In our case, it is fixed to 6 Gflops. Links are characterized by
the bandwidth and latency, which are fixed to 10 Gbps and one usec, respectively.
Bandwidth was intentionally set to a moderate value, since each simulated scenario has a limited duration.
High bandwidth values would mask out the effect of communication cost on job completion time.
The platform description that is fed to SimGrid also lists the route for each pair of nodes.
In this way, we ensure that the topology simulated matches exactly the topology assumed for deriving the
mapping of processes to platform nodes.

\subsection{Process mappings improving application performance}
\label{sec_eval_part_a}
In this section we assess the efficiency of the Scotch library in producing process-to-node mappings that improve
performance. We compare Scotch with the following process placement approaches:
a)\textit{default-slurm}, b)\textit{random}, and c)\textit{greedy}.

The \textit{default-slurm} approach refers to the default policy employed by Slurm.
The \textit{random} scheme randomly picks the node on which each process will be assigned.
Finally, the \textit{Greedy} placement approach, sorts all
different process pairs in terms of total traffic exchanged. Then, it
iterates over all pairs, starting from the one with the higher volume.
The goal is to place the processes involved as close as possible starting from a
distance of one hop.

For the evaluation purposes of this section we rely on two benchmarks:
LAMMPS \cite{ref_url_lammps_bench} and NPB-DT from the NAS parallel benchmark suite~\cite{Bailey91thenas}.
For NPB-DT we focus on the class C variant that involves $85$ processes.
LAMMPS is a state-of-the-art molecular dynamics code, while DT falls in the subcategory
of unstructured computation, parallel I/O, and data movement.

The rationale for selecting LAMMPS and NPB-DT for the evaluation is to capture three key parameters that affect the
performance of similar resource allocation approaches: the communication to computation ratio, the mix of point-to-point
and collective communication, and the communication pattern. Our approach minimizes the communication cost;
thus in applications where computation outweighs the communication, the expected speedup is insignificant.
Both benchmarks spend a significant fraction of their execution time for communication.
Additionally, in collective communication there are no specific pairs with remarkably more traffic so there is little
room for minimizing the communication.
NPB-DT is dominated by point-to-point traffic while LAMMPS exhibits a significant amount of collective traffic.

These two benchmarks exhibit different communication patterns.
As also shown in Figure~\ref{fig_lammps_heatmap}, LAMMPS exhibits a more regular communication pattern. Data
points in the corresponding traffic heatmap are arranged on a uniform manner around the main diagonal.
Each process in LAMMPS $i$ mostly communicates with processes that lie in the range $[i-k, i+k]$ for some small
value of $k$.
On the other hand, as shown in Figure~\ref{fig_npb_dt_c_heatmap}, NPB-DT exhibits a more irregular communication
pattern with no traffic around the main diagonal.
The main reason for using these two benchmarks is to challenge the assumptions of Scotch and \textit{default-slurm}.
Slurm's allocation policy iterates over the available nodes in a sequential manner.
As a result, it is highly probable for processes with ranks in some range $[i-k, i+k]$ to be placed on topologically
nearby nodes. As far as quality of mappings produced by Scotch is concerned, the work in~\cite{8411068} has outlined that in some cases, they might introduce performance degradation. On the other hand, the default placement policy used by Slurm is not expected to perform well on an irregular communication pattern like the one exhibited by NPB-DT.

Simulated runs of the aforementioned benchmarks are performed as follows: First, the assignment of processes to
platform nodes is derived. Then, this mapping is fed to SimGrid in the form of a machine file.
For all the simulated results in this section, the platform assumed is an 8x8x8 Torus, i.e. 3D Torus with 8 nodes on
each dimension.
Using SimGrid's smpirun tool, we run a simulation of the corresponding application. For NPB-DT, the performance metric
is completion time. For LAMMPS, the number of timesteps per second (reported by the application) is used as the
performance metric.  For the case of Scotch, the application's communication
graph is extracted through a trial run with the profiling tool described in Section~\ref{tofa_approach}.
Then, this communication graph is given as input to Scotch, along with a representation of the platform.
Scotch in turn produces the mapping of that job's processes on platform nodes. Finally, this mapping is given
as input to SimGrid.

\begin{figure}
     \centering
     \subfloat[]
     {
         \includegraphics[scale=.37]{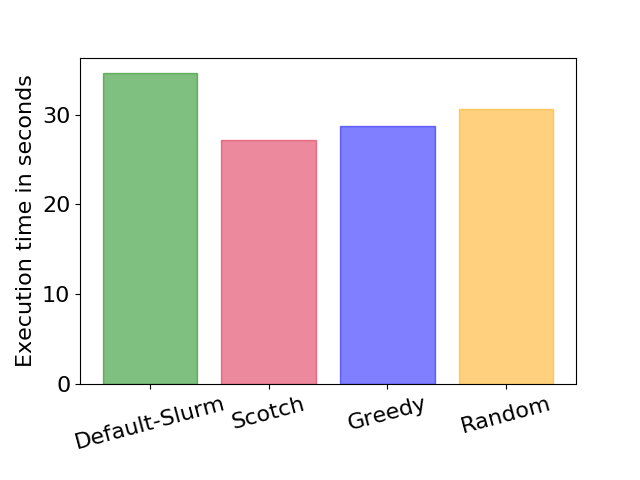}

         \label{fig_eval_npb_dt_runtime_no_error}
     }
     \subfloat[]
     {
         \includegraphics[scale=.37]{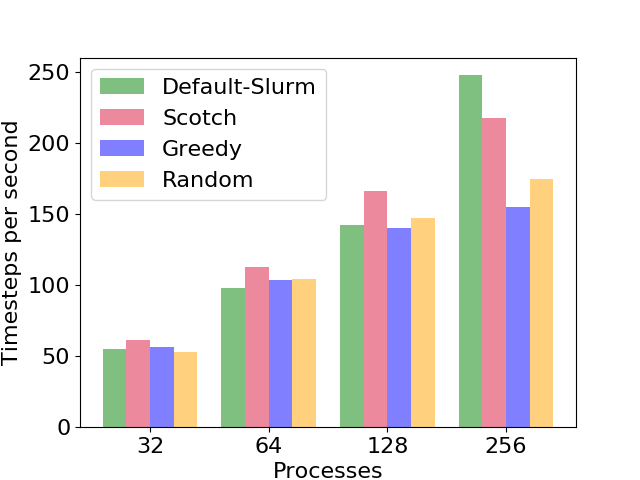}
         \label{fig_eval_lammps_runtime_no_error}
     }
     \caption{a)Execution Time for NPB-DT, b)Timesteps/s simulated for LAMMPS}
\end{figure}

As Figure~\ref{fig_eval_npb_dt_runtime_no_error} shows, for the NPB-DT benchmark, Scotch achieves the lower
execution time, with the greedy heuristic coming next. The execution time achieved by Scotch is $22\%$, $3\%$,
and $11\%$ lower than Default-slurm, Greedy, and Random, respectively.
Figure~\ref{fig_eval_lammps_runtime_no_error} depicts the timesteps per second achieved, by the various process
placement approaches for the case of LAMMPS, with different number of processes.
Higher values correspond to higher throughput in terms of simulated timesteps per second, and thus to better performance.
For the case of LAMMPS with 32, 64 and 128 processes, Scotch
outperforms all other placement approaches. For the case of LAMMPS with 256 processes though, the default placement
policy of Slurm achieves higher performance. As discussed in the beginning of this section, the default placement
policy of Slurm is expected to be beneficial for applications exhibiting a more regular communication pattern like
LAMMPS. Another reason for this performance difference is the arrangement of the underlying topology.
To further explore its effect,
different instances of the 3D torus topology are generated (all with a total of 256 nodes):
8x8x8, 4x8x16, 8x4x16, 4x4x32, 4x32x4.  For each topology arrangement, we derive an assignment of processes to platform
nodes, both through Scotch and the default placement policy of Slurm. This assignment is then provided to
SimGrid in the form of a machine file before simulating the execution of LAMMPS.

\begin{table}
  \centering
\caption{LAMMPS timesteps/s for different 3D torus topology arrangements}\label{tab1}
\begin{tabular}{|l|l|l|}
\hline
Topology arrangement & Default-Slurm & TOFA \\
\hline
8x8x8 & 247.5 & 217.2\\
4x8x16 &  188.9 & 210.1 \\
8x4x16 & 232.3 & 240.9\\
4x4x32 & 212.8 & 242.3\\
4x32x4 & 159.0 & 207.4\\
\hline
\end{tabular}
\end{table}
The corresponding results are summarized in Table~\ref{tab1}. There is significant
variability in the performance achieved by both approaches, with TOFA being less sensitive to the topology arrangement
than Default-Slurm. A time-based breakdown per MPI primitive reveals that the topology arrangement may have an adverse
effect on the time spent in collective operations. For the case of the 4x8x16 arrangement for example, the average
process time spent in broadcast is significantly inflated for the case of Default-Slurm.

\subsection{Process mappings improving resilience in the presence of node Failures}
\label{sec_eval_part_b}
In this section we evaluate the performance of the proposed approach in the presence of node failures.
Instead of simulating a single MPI job instance,
we use {\em job batches}, each consisting of 100 instances of the same MPI application.
We use two criteria to evaluate each process placement approach: batch completion time, and abort ratio.
The batch completion time is the total time required to complete the queue of 100 instances.
The abort ratio is the fraction of instances that were aborted due to one or more node outages.
As explained in Section~\ref{tofa_approach}, when a job fails we assume that it is restarted from scratch,
rather than being restored from a checkpoint.
This assumption simplifies the way we update batch completion times in the presence of
node failures. Each time a job is aborted, the batch completion time is augmented by a time interval
equal to a successful run, and then the job is restarted.

The evaluation experiments in this section compare Slurm's default placement approach (Default-Slurm)
with the proposed placement approach (TOFA).
Different simulated scenarios are examined based on the following parameters:
\begin{itemize}
    \item MPI application
    \item $N$: Number of MPI processes involved
    \item $p_{f}$: node outage probability
    \item $n_{f}$: number of nodes emulated in the failed state
    \item $N_{f}$: set of nodes emulated in the failed state
\end{itemize}
The MPI applications simulated are LAMMPS (64 processes) and NPB-DT (85 processes) of class C.
For each application, 10 different batches are created consisting of 100 instances each.
For each batch, the nodes to populate set $N_{f}$ are randomly selected and remain the same for
all 100 instances of the same batch.
Similarly to the evaluation results presented in Section~\ref{sec_eval_part_a},
the platform assumed consists of $512$ nodes arranged in an 8x8x8 Torus.
In each SimGrid scenario, for emulating a node as being in the failed state, the following approach is
followed.
Each node from set $N_{f}$ is assigned a fixed outage probability $p_{f}$ which is the same for
all nodes. Based on this probability, we determine whether the node will be emulated as being in
the failed state, or not, in each scenario.
This means that for each simulated scenario, a different subset of nodes in $N_{f}$ will
be emulated as being in the failed state.
SimGrid allows to specify different values for a specific link's capacity, at different points in simulated time.
Specifying a value of zero, will result in all transmissions in a link failing, and consequently, in abortion of the MPI
program simulated.
These variations in link capacity are defined in the platform description that is provided as
input to SimGrid. Thus, for every node that will be emulated as being in the failed state, the platform description
is updated by assigning a zero bandwidth value to all links that this node participates.

\begin{figure}
  \centering
\includegraphics[width=0.85\textwidth]{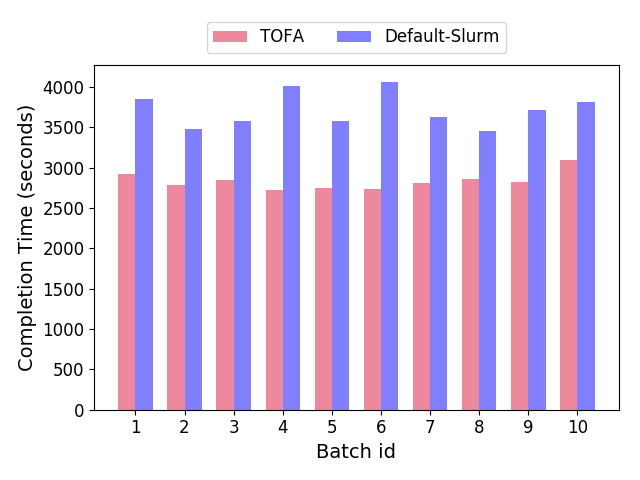}
\caption{NPB-DT batches completion time, 16 faulty nodes with 2\% chance each} \label{fig_DT_faults}
\end{figure}

Figure~\ref{fig_DT_faults} depicts the completion time for 10 different batches of NPB-DT jobs with 85 processes each.
For each batch, 16 nodes out of all 512 are randomly selected and are assigned an outage probability of $2\%$.
As this figure shows, for all batches, TOFA
achieves significantly lower completion time than the default placement policy of Slurm.
More precisely,
the average improvement in batch completion time, over all 10 batches, is
$31\%$.
This drop in batch completion time is due to two factors: (1) the reduction in the
communication cost as the result of topology- and application profile-aware placement;
and (2) the drop in the jobs being aborted due to node outages.
The average job abort ratio (over 1000 simulated NPB-DT instances) is $2\%$ for the case of TOFA, and $7.4\%$
for Slurm's default process placement policy.
The average duration of each simulated scenario reported in Figure~\ref{fig_DT_faults} is
limited to $30$ seconds (on average). The effect of aborted jobs on average
batch completion time is expected to be even more significant for larger problem sizes.

\begin{figure}
     \centering
     \subfloat[]
     {
         \includegraphics[scale=.37]{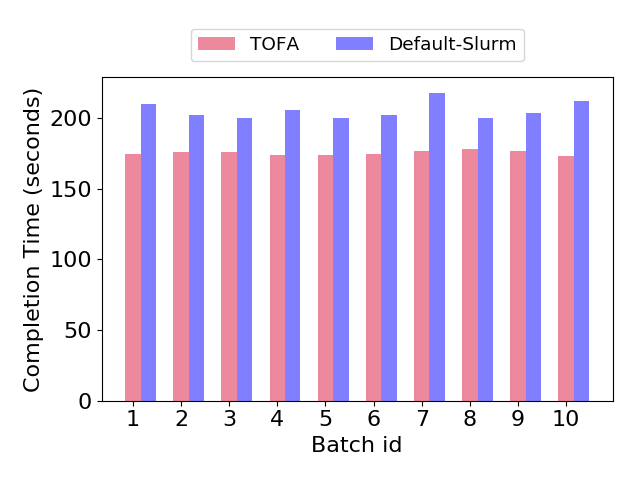}

         \label{fig_lammps_errors_8nodes_2percent}
     }
     \subfloat[]
     {
       \includegraphics[scale=.37]{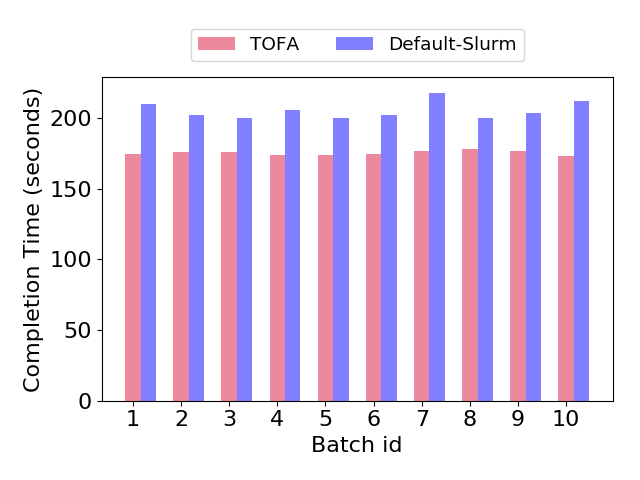}
         \label{fig_lammps_errors_16nodes_2percent}
     }
     \caption{a)LAMMPS, 8 faulty nodes 2\%, b)LAMMPS, 16 faulty nodes 2\%}
   \end{figure}

The following simulation results are derived from LAMMPS runs of the {\em rhodopsin} problem with 64 processes.
Two different scenarios are simulated. In the first one, eight nodes have a $2\%$ probability of
entering the failed state. In the second one, the corresponding number is 16, therefore approximately $3\%$ of
total nodes have an outage probability of $2\%$.
Figures~\ref{fig_lammps_errors_8nodes_2percent} and \ref{fig_lammps_errors_16nodes_2percent} depict the
completion time for 10 different batches of the LAMMPS application, for the two aforementioned scenarios,
and two different process placement approaches: \textit{TOFA} vs. Slurm's default placement policy
(\textit{Default-Slurm}). For all 10 different batches, the completion time achieved by \textit{TOFA} is lower.
On average over all 10 batches, \textit{TOFA} achieves $17.5\%$ and $18.9\%$ lower batch completion time than
\textit{Default-Slurm}, for the cases of $8$ and $16$ failing nodes, respectively.
As it is expected, when the fraction of nodes that are probable to enter the failed state increases, the gain of
allocating resources taking into account these probabilities also increases.
As in the case of NPB-DT, the drop in the batch completion time is attributed to both the reduction of the
communication cost and the reduced overhead of failed jobs that need to be restarted.
Another interesting observation is that, for the scenario depicted in Figure~\ref{fig_lammps_errors_8nodes_2percent},
where $8$ nodes are emulated as
faulty, \textit{TOFA} always manages to find 64 consecutive non-faulty nodes and thus, achieves a zero job abort
ratio. This is not the case though, for the scenario with 16 faulty nodes, where the corresponding job abort
ratio achieved is $1.1\%$ for \textit{TOFA} and $4.0\%$ for \textit{Default-Slurm}.

It is also interesting to compare the average benefit over all 10 batches achieved by \textit{TOFA}
for NPB-DT and LAMMPS. As also discussed in Section~\ref{sec_eval_part_a}, LAMMPS exhibits a more regular communication pattern
compared to NPB-DT. As a result, the benefit of a topology aware process placement approach over the default placement
policy of Slurm is expected to be lower. Indeed, the corresponding benefit in average batch completion time achieved
by \textit{TOFA} is $18.9\%$ for the case of LAMMPS and $31\%$ for the case of NPB-DT (with 16 failing nodes).

\section{Conclusions} \label{sec_conc_future_work}
In this work, we present a process placement approach for MPI jobs that improves completion time.
For deriving the assignment of job processes to nodes, we take into account both the topology and the job's
communication patterns. Differing from similar approaches, we also take into account node failures for
post-processing the graph that models the topology. The goal of this approach is to reduce communication cost
due to inter-node traffic and also reduce the overhead of restarting jobs that were aborted due to node failures.
The assignment of processes to nodes is formulated as a topology mapping problem.
TOFA has been integrated into Slurm, using plugins that extend its functionality.
The proposed approach has been evaluated in the SimGrid environment using two different benchmarks.
For the case where around $3\%$ of all nodes have an outage probability of $2\%$, TOFA reduces completion
time by $18.9\%$, for jobs with a regular communication pattern. The corresponding reduction for the case of a
benchmark with a more irregular pattern is $31\%$. The proposed approach also achieves a notable reduction in
the probability of a job being aborted due to node failures.

Part of our ongoing work explores parameters affecting the efficiency of the proposed process placement
approach. We consider the algorithms used to implement each collective operation, the arrangement and dimension
of the available platform, and finally, the metric used to express communication patterns. Part of our future
work is to evaluate different policies for capturing the effect of node outages on the cost of traversing a path,
and gaining insight into the trade-off between a job's completion time and probability of abort.

\section*{Acknowledgments}
We thankfully acknowledge the support of the European Commission under the Horizon 2020 Framework Programme for Research and Innovation through the projects EuroEXA (Grant Agreement ID: 754337) and ExaNeSt (Grant Agreement ID: 671553).

\bibliographystyle{splncs04}
\bibliography{isc2019bib}

\end{document}